# Novel community data in ecology- properties and prospects

Florian Hartig[1,*], Nerea Abrego[2], Alex Bush[3], Jonathan M. Chase[4], Gurutzeta Guillera-Arroita[5], Mathew A. Leibold[6], Otso Ovaskainen[2,7], Loïc Pellissier[8,9], Maximilian Pichler[1], Giovanni Poggiato[10], Laura Pollock[11], Sara Si-Moussi[10], Wilfried Thuiller[10], Duarte S. Viana[12], David Warton[13], Damaris Zurell[14], Douglas W. Yu[15, 16]

1) Theoretical Ecology, University of Regensburg, Regensburg, Germany
2) Department of Biological and Environmental Science, University of Jyväskylä, P.O. Box 35 (Survontie 9C), FI-40014 Jyväskylä, Finland
3) Lancaster Environment Centre, Lancaster University, UK
4) German Centre for Integrative Biodiversity Research (iDiv) Halle-Jena-Leipzig
5) Pyrenean Institute of Ecology, CSIC, Spain
6) Biology Department, University of Florida, Gainesville, Florida, USA
7) Organismal and Evolutionary Biology Research Programme, Faculty of Biological and Environmental Sciences, University of Helsinki, P.O. Box 65, Helsinki 00014, Finland
8) Ecosystems and Landscape Evolution, Institute of Terrestrial Ecosystems, Department of Environmental Systems Science, ETH Zürich, 8092 Zurich, Switzerland
9) Unit of Land Change Science, Swiss Federal Research Institute for Forest, Snow and Landscape Research (WSL), 8903 Birmensdorf, Switzerland
10) Univ. Grenoble Alpes, Univ. Savoie Mont Blanc, CNRS, LECA, F38000, Grenoble, France
11) Department of Biology, McGill University, Montreal, Quebec, Canada
12) Doñana Biological Station (EBD-CSIC), Spain
13) University of New South Wales, Australia
14) University of Potsdam, Potsdam, Germany
15) Kunming Institute of Zoology; Yunnan, China
16) University of East Anglia, Norfolk, United Kingdom

* corresponding author, florian.hartig@ur.de

## Abstract

New technologies for acquiring biological information such as eDNA, acoustic or optical sensors, make it possible to generate spatial community observations at unprecedented scales. The potential of these novel community data to standardize community observations at high spatial, temporal, and taxonomic resolution and at large spatial scale ('many rows and many columns') has been widely discussed, but so far, there has been little integration of these data with ecological models and theory. Here, we review these developments and highlight emerging solutions, focusing on statistical methods for analyzing novel community data, in particular joint species distribution models; the new ecological questions that can be answered with these data; and the potential implications of these developments for policy and conservation.

# Introduction

Understanding the factors that govern the distribution of Earth's biodiversity across space and time remains one of the most important and challenging problems in biodiversity science. While human activities are rapidly altering the structure of biodiversity and the services it provides to humans [1], our ability to monitor, model and predict these changes have been hampered by the fact that classical methods for biodiversity monitoring are limited in spatial, temporal, and taxonomic scale and resolution, and are poorly standardized or structured [2].

In recent years, major technological innovations in biodiversity monitoring have occurred. We focus particularly on **environmental DNA** (eDNA, see Glossary), but also refer to **passive acoustic monitoring** [3–5] and visual sensors which [6], coupled with appropriate machine learning or deep learning **pipelines** [7,8], are moving the field "towards the fully automated monitoring of ecological communities" [9], promising to transform the way data on species distribution and abundance are generated for the rest of the 21st century [10]. The efficiency gains are such that hundreds or even thousands of species can be routinely detected and potentially quantified in their abundance across entire landscapes. Because community data are typically represented as rows of observations and columns of species in a **community matrix**, these are 'many-row, many-column' multivariate abundance data. We hereafter refer to them as **novel community data** (see also [10]).

The emergence of novel community data presents ecologists with many opportunities as well as challenges (e.g. [11–13]). These datasets are larger and richer in information, but they also have complicated properties such as higher rates of false positives or unreliable between-species relative-abundance information [14,15]. Novel community data therefore require appropriate statistical tools that can exploit the additional information, including community dynamics, while accounting for the added complications [16].

The sensors and technologies used to generate novel community data have been extensively reviewed elsewhere [9–11,17–23]. In this review, we therefore cover this topic only briefly and instead focus on how the combination of novel community data with new statistical tools and questions both compels and enables us to transform data analysis, expand our scientific reach, and improve biodiversity conservation and management.

## What makes novel community data really novel?

Over the past two decades, two types of data sources have dominated ecological research on species distributions, community assembly, extinction risk, and many related questions.

The first is large databases of spatial occurrence data (e.g. GBIF), range maps (e.g. IUCN), or taxa-specific monitoring schemes. These data are frequently used in **species distribution models** (SDMs, e.g. [24,25]) to estimate species' environmental niches, project future distributions under climate or land use change, or generate biodiversity metrics for conservation and management. Although widely used, these often opportunistically generated data pose important limitations for biological analysis due to their uncertain observation errors and intensities [26], and the fact that these data rarely include inventories of entire communities across trophic groups which limits their potential for understanding interactions in communities as well as ecosystems dynamics.

A second type of data is community inventories conducted for conservation and management purposes, as well as studies in metacommunity ecology [27]. The distinguishing feature of this data is that they provide both presence and (somewhat reliable) absence or abundance/biomass information for entire local communities. Such community datasets are often limited in sample size, spatial and temporal extent as well as taxonomic coverage and resolution (see [19], but see [28]).

The emergence of novel community data (Fig. 1) promises to fundamentally alter this established landscape of biodiversity observations. It is tempting to dismiss our ability to sequence environmental DNA (eDNA), ancient DNA (aDNA), and bulk-sample DNA [19,20,23,29] as well as monitor species using camera traps or passive acoustic monitoring merely as a convenient way for generating more data (i.e. big data) of the same kind that we have been collecting. Such a view neglects, however, the other dimensions on which these data differ from existing community data in terms of structure, resolution, and **metadata** acquisition:

## Structure and standardization

Especially as technology evolves and pipelines are shared, compared, and converge to common standards, community datasets have the potential to be more structured and standardized than traditional sampling schemes. They are typically generated according to a fixed plan using low-expertise collection methods (e.g. soil samples), positive and negative controls, and a standardized processing pipeline for species identification. Importantly, by standardized, we do not mean error-free. eDNA data, for example, can have considerable errors (see Box 1). However, when these errors are more homogenous and predictable, they could more easily be corrected using statistical methods than the heterogeneous errors associated with human observers.

## Spatial, temporal, and taxonomic resolution

The automated way in which novel community data are generated is more easily scaled up to achieve high spatial, temporal, and taxonomic resolution [29,30]. Different sensors will have different strengths along these dimensions, but these strengths can be combined through a combination of sensor types (see also [13]). For example, while eDNA data have particular strengths in detection sensitivity, taxonomic breadth, and community completeness (Box 1), acoustic and visual sensors are better at creating continuous community time series. Indeed, time-lapse, automatic detection or continuous recording offer the unique opportunity to capture biodiversity along the daily, seasonal or decadal time scales, something difficult to achieve with more traditional sampling schemes. An obvious way forward would be to use statistical methods to combine observations from these different data streams into a combined spatiotemporal data product (cf. [12,19], see also outstanding questions). Advanced methods may allow us to go beyond presence / absence and also identify abundance change (see [31] and Box 1), as well as genetic diversity within and between species [32]. For example, in particular for eDNA, but also for visual and acoustic sensors, it is now possible to identify entities below the species level, e.g. **exact-sequence variants (ESVs)**. Next-generation methods may also allow individual-level identification and tracking (via genetic data and image-analysis) to estimate abundance or dispersal distances and patterns.

## Metadata acquisition and matching to other data sources

Assuming that novel community data acquisition is more technologically driven so that location and time stamp information is always available, it should be easier to complement it with meta-data such as the specific climatic conditions of the observation. These can then be matched with other data products such as remote sensing products, phylogenetic or trait information, or direct **biotic interactions** extracted from visual, acoustic data or eDNA analysis [33]. Such matched data products could form the basis of the essential biodiversity variables of the GEO-BON platform [34].

## Challenges to data quality

Parallel to these advantages, there are also persistent challenges to all major novel community data streams. Most importantly, ecologists have often been skeptical about the quality of observations derived from novel community data pipelines. We acknowledge that each sensor type faces certain technical challenges, some inherent in the measurement process itself (sensitivity, field of view) and some inherent in the analysis pipeline (incomplete **DNA barcoding** reference databases,

PCR errors, transferability and generalizability in deep learning methods used for species recognition). The combination of measurements and pipeline can introduce errors and biases that need to be addressed (see Box 1 for a discussion of the eDNA pipeline). However, as we point out in our section on statistical analyses, once pipelines are becoming more standardized and protocols are shared, there is the potential to create parameterized statistical models to account for these errors in the statistical analysis.

# Using novel community data to answer old ecological questions

Having established that novel community data will provide not only a larger sample size, but a much richer, standardized, and interconnected data product than traditional biodiversity monitoring data, we focus on the way these data will transform the way we can approach classical and new ecological questions. We will divide this discussion into five themes: i) **species associations**, ii) biotic, especially trophic interactions, iii) beyond the species concept, iv) real-time monitoring and long time series, v) understanding ecosystems as complex systems.

## Species associations

Because novel community data provide relatively complete community inventories, they are well suited to measure species associations. Species associations can arise through measured environmental factors and/or spatial patterning, but even when this is accounted for (see section "statistical tools"), species often show remaining associations. These may be artifacts due to unmeasured (or inadequately) measured environmental factors or to inadequately quantified spatial patterning [e.g., 35–37], but they can also reflect species interactions. Being able to comprehensively quantify species associations, especially when used in conjunction with direct observations of biotic interactions (see below), we have the potential to make progress on the old question of disentangling space, abiotic and biotic factors as drivers of (meta)community assembly [38–40]. Moreover, if the data contains both spatial and temporal dimensions, associations can be investigated both over time and space, lending more confidence that these associations are no statistical artifacts [41]. Finally, even if the causes for spatial associations cannot be resolved, they reduce the unexplainable variation in the community composition and may thus give a more realistic evaluation of stochasticity in community dynamics and assembly rules (e.g. [39]).

## Biotic interactions

Novel community data, particularly eDNA data, can also be used to directly infer species interactions, both trophic and mutualistic [42]. The most straightforward way to obtain trophic interactions and thus infer whole food webs is to analyze the gut contents of individuals to infer diet (see, for example, [43], who analyzed the gut content of coral reef fish to establish a complex marine food web). It is, however, also possible to infer host-vector-pathogen networks [44] or mutualistic interaction networks using interaction residues, e.g. by analyzing pollen [45] or eDNA traces on flowers or pollinators [33]. An interesting avenue is to link such eDNA data with background knowledge from observations or experiments [46] and link those data to disturbances [47] or ecosystem functioning [48].

## Beyond the species concept

Another area where novel community data may lead to progress is by challenging the central role of the species as the basic unit for quantifying biodiversity and community patterns. While we believe that the species concept remains central to ecology, novel community data may increase taxonomic resolution to the subspecies or even ESV level, thus not only solving the problem of cryptic species [49] but also revealing intraspecific patterns of reproductive isolation as well as interspecific gene flow on a large scale (e.g. 'macrogenetic', cf. [50,51]). An important question is how such a more gradual species concept could be mixed with other concepts such as competition, distribution, niche, or extinction, which are central to both ecology and practical conservation [52].

## Real-time monitoring, nowcasting and ancient DNA

An advantage of acoustic and visual sensors over eDNA is their high temporal resolution, which offers the potential to observe short-term changes in population size, species interactions or habitat preferences, or phenological changes. However, also eDNA is well suited for generating time series (e.g., [20], Fig. 2). Both offers the potential for real-time monitoring and nowcasting of biodiversity changes, biological invasions and pathogen outbreaks [53,54]. Another interesting idea is the ability to generate observations and time series from the past using ancient DNA [20,55], which could be instrumental in understanding human impacts on ecosystems in the Anthropocene.

**Ecosystems as complex systems**

Finally, the fact that novel community data provide direct measurements of species interactions (i.e. the trophic structure) together with community presences or abundances at high spatiotemporal resolution may finally deliver the data needed to for the old aspiration of "modelling all life on earth" [56], i.e. understanding ecosystems holistically as complex systems and describing their various interactions through mechanistic ecosystem or macroevolutionary models (e.g., [57]).

# Statistical tools for novel community data

The "law of the instrument" famously warns us that "if all you have is a hammer, everything looks like a nail" [58]. The saying alerts us to the fact that, to use the full potential of novel community data, we require analytical tools that make use of the particular properties of the data and connect them to ecological theory. We see three main directions in which statistical methods for novel community data should be developed: community and metacommunity analysis, time series analysis, and network analysis.

## Community and metacommunity analysis

Community and metacommunity analysis aims to understand how community composition changes as a function of the environment and possibly interactions between communities. Statistically, we can approach this problem from at least three angles: we can use differences or changes in community composition as a response (e.g. ordination, Mantel tests or regressions on distance matrices [59]), we can use constrained ordinations to partition effects on community composition between space and environmental predictors, or we develop statistical models that directly predict community composition (as done, for example, in joint species distribution models (jSDMs), see [60–62]. While each of these approaches has its strengths, we find the option to model communities directly particularly attractive because it infers species-specific environmental preferences, spatial effects and species associations (for details, see Box 2), all quantities that are biologically interpretable and useful for making predictions.

## Time series analysis

A limitation of static data is the difficulty of distinguishing correlation from causation as well as directional effects between species (e.g., hierarchical competition). Temporal data provides the potential to alleviate some of these problems. A prominent idea in causal time series analysis is the concept of Granger causality [63], which posits that because the cause must precede the effect, we can regress our observations (in this case the community composition at each time step) against the observations of previous time steps. In theory, this approach could also infer asymmetric interactions, and it has been argued that interactions based on such a temporal or spatio-temporal approach are more suitable for inferring true biotic interactions (see [64] and Fig. 2, for an implementation in an extended jSDM).

## Network analysis

A third avenue for statistical analysis is the comparison of co-occurrence networks inferred through the analysis described above, and the mutualistic, trophic or competitive biotic interaction networks that are generated, for example, by sequencing gut content (see also Fig. 1). This research direction could leverage methods from the field of network analysis [65], which often struggles with the same data limitations as in community ecology. Novel community data could allow us to analyze larger and more complex networks (e.g., [66]), analyze how these networks change across environmental gradients [67], and link those patterns to community data to understand how biotic interactions, in conjunction with environment and space, give rise to spatio-temporal biodiversity patterns [68]. For example, measured associations between species may change with scale [69], but it is unclear how those changes reflect biotic interactions and it is also possible that two species are connected by an interaction (e.g., predator-prey), but not show any spatial association [70].

## Statistical observation models

In the above-mentioned approaches, it will likely be crucial to include statistical observation models that correct for detection probabilities and taxonomic misclassification (e.g. [71]). The problem of including detection probabilities is not specific to novel community data (e.g. [72,73]), but detection errors may be more pronounced in novel community data (e.g. Box 1). On the positive side, due to standardized pipelines, detection and measurement errors may be easier to estimate, and certain detection biases (e.g. difficulty estimating relative abundances due to species effects, Box 1) may be irrelevant for certain types of statistical analyses such as species distribution modeling. First attempts to combine community models with observation or occupancy models to produce corrections for detection errors [74–76] already exist. A challenge for the future is to make them broadly available and fit for the computational demands of large datasets.

## Improving predictions

An outstanding question is how useful novel community data is to create predictions, in particular under various global changes. Community models (e.g. Box 2) that include species associations do not necessarily predict better unless we already have information on some of the species [37,77]. One could hope, however, that predictive advantages may be created from temporal or spatio-temporal community data, because in such data, causal effects and directional interactions can be better identified [63], and it is possible that machine learning and deep learning methods [8] are better able to extract useful information out of these complicated datasets.

# Leveraging novel community data to achieve socio-ecological resilience

Beyond scientific progress, novel community data may also enhance society's ability to achieve *socio-ecological resilience*, which is the ability of human institutions to strengthen ecosystem resistance to and recovery from undesirable change [78]. In their seminal paper, Dietz *et al*. [79] describe five elements of socio-ecological resilience: (1) knowledge generation, (2) providing infrastructure, (3) political bargaining, (4) enforcement, (5) institutional design, and adaptive governance.

The most obvious avenue for novel community data to contribute to these elements is our ability to generate *high-quality*, *granular*, and *timely* knowledge about ecosystem status, health and change, uncertainty levels, values, and the magnitude and direction of anthropogenic impacts. In addition, as methods become more automated, independent parties can collect and compare large biodiversity datasets, which will make this knowledge more *trustworthy* and *understandable* [80]. This information can in turn make political conflicts more resolvable and enforcement more effective.

Governments can encourage this process by 'technology forcing' and use the advantages of novel community data to redesign environmental institutions [81]. An example of such a process is the UK's District Licensing offset market for the great crested newt (Box 3). Also, both governmental and non-governmental institutions have an opportunity to redesign scientific institutions leverage the advantages of novel community data. For instance, although most regulatory use of eDNA is still only single-species detection [81], in the US, these data are being combined into a multi-species database, the Aquatic eDNAtlas Project. To facilitate such a process, strict sampling protocols, reference datasets and pipelines for creating biodiversity data (e.g. AI models for species recognition, barcode databases) should be freely available and integrated into global monitoring schemes and databases such as GBIF, IUCN, and GEOBON (e.g. [21,82]). Based on these, one could create policy-relevant data products such as global biodiversity intactness maps with granular and timely data (e.g. STAR, see [83]). Bayesian optimal design methods could be used to identify data gaps and thus to prioritize funding for initiatives to address those gaps. For industry, the availability of such data may help to include ecological impacts into corporate decision making. For example, the Task Force on Nature-Related Financial Disclosures (TNFD, [tnfd.global](tnfd.global)) has developed an analysis framework to assess corporate exposure to nature-related risks and opportunities.

# Concluding remarks: Outlook for ecological research

In conclusion, we believe that novel community data offer new opportunities for the analysis of biodiversity. Novel community data can provide spatiotemporal information on community abundance with high spatial, temporal, and taxonomic resolution, in conjunction with traits, standard abiotic predictors, and partly observed true biotic (mutualistic and trophic) interactions. While the need and value of multi-faceted biodiversity analyses has been acknowledged for some time, the main reason for this development is the emergence of sensors that inherently produce community rather than single-species data. The lower costs and wide availability of these data have important implications and open new avenues for testing ecological concepts and theories.

We believe that (joint) species distribution models, network analyses and related statistical tools inherited from causal analyses could act as some of the core analytical tools to map these data to important ecological concepts and theories, in particular niche theory, metacommunity theory, and network theory. Beyond this, novel community data also have high potential to provide crucial information for environmental management and biodiversity conservation.

Challenges for the future (see Outstanding questions) are the creation of suitable data products, which includes comparable field designs and bringing together existing data in common databases, the establishment of accessible statistical pipelines, and, connected to that, the link to ecological theory and actionable predictions for management and conservation.

# Acknowledgements

This paper originated from several meetings of the sCom working group supported by sDiv, the Synthesis Centre of iDiv (DFG FZT 118, 202548816). We thank the editor and two anonymous reviewers for their helpful comments on earlier drafts of this manuscript, and Andrew Buxton and Sarah Garratt from NatureSpace Partnership for comments on the newt offset market. WT and GP also acknowledge support from the European Union's Horizon Europe under the project NaturaConnect (grant agreement No 101060429) and OBSGESSION, as well as the French Agence Nationale de la Recherche (ANR) through the Gambas project (ANR-18-CE02-0025). SS-M and WT were funded by the French Biodiversity Office through the FrenchBiodiv project. OO was funded by Academy of Finland (grant no. 336212 and 345110) and the European Research Council (ERC) under the European Union's Horizon 2020 research and innovation programme (grant agreement No 856506; ERC-synergy project LIFEPLAN). GGA is currently supported by a ´Ramón y Cajal´ grant (RYC2020-028826-I) funded by the Spanish Ministry of Science and Innovation, the Agencia Estatal de Investigación (10.13039/501100011033) and "ESF Investing in your future". MAL was funded by U.D. NSF award 2025118. DSV is currently supported by a ´Juan de la Cierva Incorporación´ postdoctoral grant (IJC2020-044545-I) funded by the Spanish Ministry of Science and Innovation. DWY was supported by the CAS Strategic Priority Research Program (XDA20050202) and the CAS Key Research Program of Frontier Sciences (QYZDY-SSW-SMC024).

## Glossary

**Biotic interaction:** a direct (e.g. competitive, mutualistic, trophic) interaction between individuals of two different taxa

**Community matrix:** a matrix of community observations, traditionally with rows = samples or sites and columns = species or taxa, which characterize presence, presence-absence, or abundance or biomass for each species / site combination.

**Cryptic species:** species that are morphologically indistinguishable but genetically distinct and reproductively isolated and can thus only reliably be identified with molecular analyses.

**Environmental DNA (eDNA):** DNA isolated from environmental samples, including both extraorganismal (trace) and organismal eDNA. For example, bulk-arthropod samples contain both organismal eDNA from arthropods and trace eDNA from vertebrates (e.g. blood, feces, skin).

**Exact-sequence variants (ESVs):** unique DNA sequences that are identified from high-throughput sequencing. Unlike more traditional operational taxonomic units (OTUs, see below), which cluster non-identical but similar sequences, ESVs describe identical nucleotide sequences.

**DNA barcoding:** species identification using a short section of DNA from a specific gene or genes, which is mapped against a barcoding reference database

**joint Species Distribution Model (jSDM):** a statistical model that describes a vector of community (multi-species) presences or abundances as a function of abiotic, biotic or spatial predictors (like an SDM) and an additional component, which consists of residual covariances between the modeled species, describing positive or negative species associations.

**Metadata:** in general, data describing other data. In the context of this paper, we include in this definition all data that complement the primary community observations. This includes location, time, and biosensor data (e.g. temperature) as well as matched environmental predictors or remote sensing products.

**Operational Taxonomic Unit (OTU):** a group of haplotypes that are clustered together based on their sequence similarity to form distinct taxonomic entities, typically species.

**Passive acoustic monitoring:** deployment of acoustic sensors in the field to detect sounds created by wildlife and the surrounding (soundscape). This data can be processed by experts or machine learning methods to classify the sounds of specific species or communities.

**Pipeline:** a series of computational and analytical steps to process and analyze raw sensor data such as sequencing data, acoustic observations, or pictures.

**Species Distribution Model (SDM):** a statistical model that relates species presence or abundance data to a set of abiotic, biotic or spatial predictors.

**Species association:** a correlation or association of occurrence, abundance, or distribution of two taxa, which can be due to biotic interactions, (missing) environmental covariates, distributional disequilibrium, and other reasons.

**Novel community data:** large community datasets generated by automated pipelines such as eDNA sequencing and electronic sensors (e.g. bioacoustics, camera traps, citizen-science app networks).

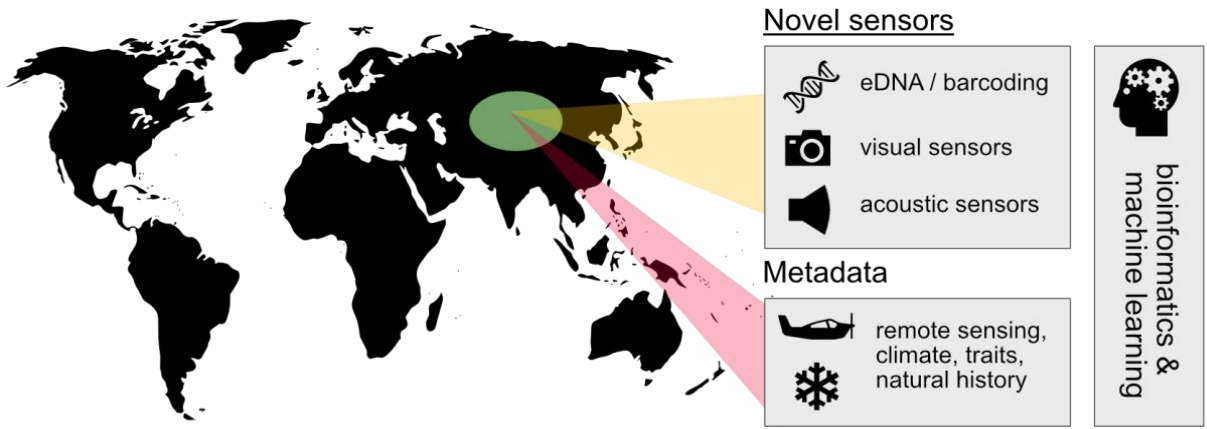
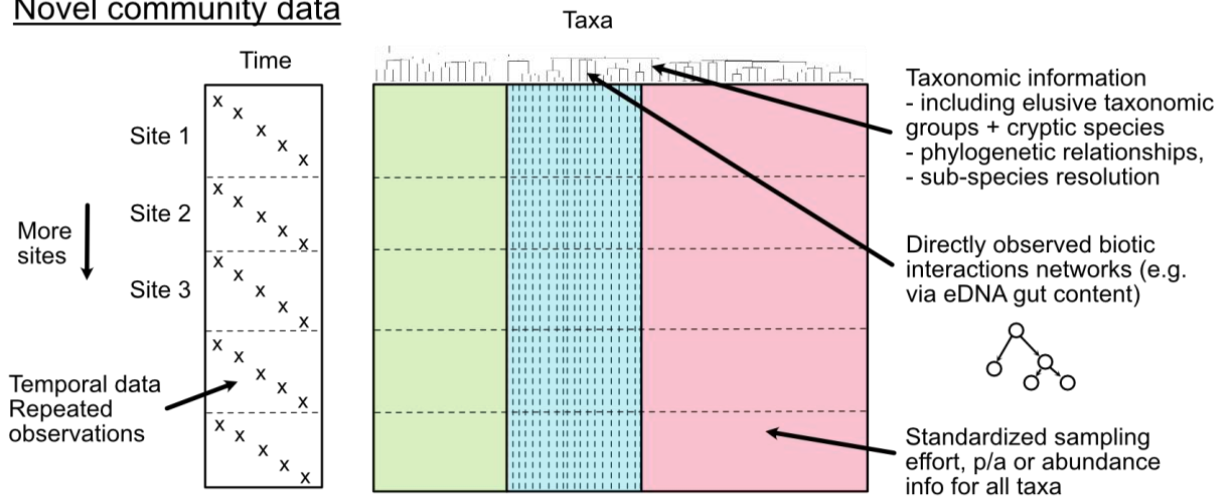
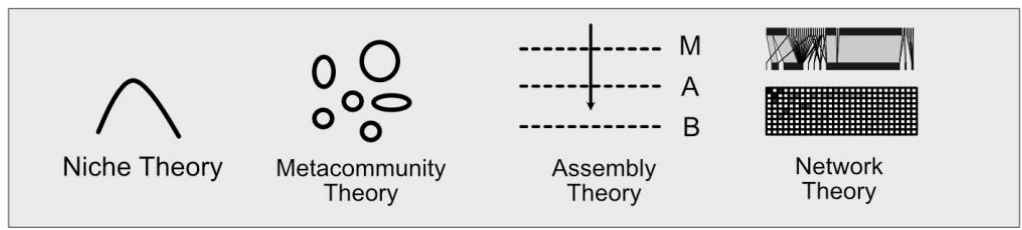

**Fig. 1:** Novel sensors including eDNA (see also Box 1), visual or acoustic sensors, linked with metadata acquisition, create detailed community inventories. If replicated in space and time, this gives rise to novel community data. Note that the novel community data structure in the center is more information-dense in many dimensions beyond spatial replicates, including time, taxonomic relationships and interaction information. The resulting opportunities to address ecological questions will be discussed in the following sections.

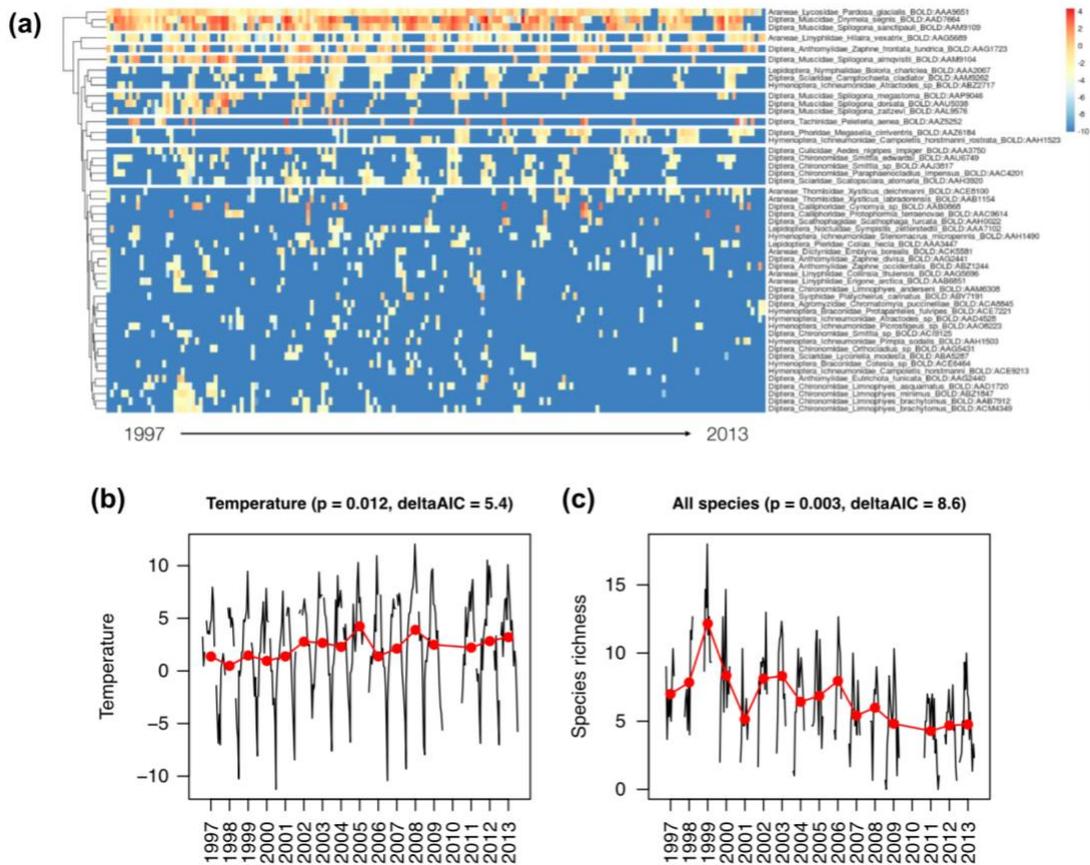

**Fig. 2:** Abrego et al. (2021) [29] analyzed a 14-year-long, weekly community time series of arthropod community dynamics from Greenland, resolved to the species level by eDNA mitogenome mapping. Panel a) shows the community matrix with species x time. During the study period, temperature increased by 2°C and arthropod species richness halved (panels b respectively c, reprinted from [29]. In their analysis of the data, the authors show that abiotic variables alone are insufficient to predict species responses, but with species interactions included, the predictive power of the models improves. Trophic cascades thereby emerge as important in structuring biodiversity response to climate change. The study emphasizes the potential of eDNA data to generate high-resolution community time series and thus to understand the complex interplay of biotic and abiotic effects in climate change impacts. The analytical tools used to arrive at these conclusions are explained in the next section (statistical tools).

## Box 1: An overview of the eDNA pipeline

All species shed DNA into the environment. We refer to this DNA isolated from environmental substrates, even air [84,85], as eDNA [23,86,87]. eDNA can either be sequenced *en masse* and processed *in-silico* to find taxonomically informative sequences (metagenomics) or read after targeted amplification of taxonomically informative sequences in the laboratory (metabarcoding). In both cases, the resulting DNA sequences, 'reads', are either compared to DNA-barcode reference databases to assign taxonomies [88] or clustered to **operational taxonomic units** (OTUs). eDNA technology thus enables the standardized detection of many species, including cryptic, difficult-to-observe, small, and low abundance species, from easily collected samples.

Challenges for this field are the currently high diversity of different bioinformatic pipelines to curate, clean and cluster eDNA sequences (but see comparative analyses to identify best-suited pipelines for particular purposes, e.g., [89]), as well as eDNA specific sampling and detection error (Table B1, see also [14,75,90,91].

For example, stochasticity in laboratory pipelines can obscure the expected positive relationship between the eDNA biomass and the resulting number of reads. By adding a DNA spike-in to each sample, the number of spike-in reads per sample can be used to recover this relationship [75,90]. Similarly, sample contamination can result in false-positive errors, but good practice constrains such events to be rare and weak, letting false positives be identified [75,92].

A further challenge with eDNA data is estimating absolute species abundances because the number of eDNA 'reads' per species depends in part on unknown, species-specific rates of release, degradation, and PCR efficiency ('species effects') [15]. However, if (1) these rates are approximately constant across study sites, and (2) pipeline stochasticity is accounted for (via spike-ins), then for each species, change across samples in the number of reads assigned to each species can be interpreted as change in each species' abundance [31,75,90,93].

Finally, taxonomic assignment error, due to incomplete reference databases and variation across species in genetic diversity, is an important problem. It can be accounted for via Bayesian algorithms that are trained to estimate the degree of sequence similarity needed to assign membership to a given rank within a given taxon [94,95].

**Table B1:** The two stages of DNA-based surveys and the sources of false-negative error, false-positive error, and row, column, and cell effects in the output sample x species table (adapted from [75]).

| Stage 1 - eDNA biomass collection | | Analogues in conventional surveys |
|---|---|---|
| Species effects | Every sample collects a certain amount of eDNA biomass of each species, which is proportional to the species' biomass available at the site. However, the proportionality constant is marker- and species-specific and is unknown, since rates of DNA release, 'catchability', and degradation differ across species and physiological states (a 'column' effect). | Species differ in their detectability by human observers or by trapping bias. |
| Noise | The amount of eDNA biomass collected per species varies stochastically among samples collected at the same site and time (a 'row' effect), including outright collection failure (false negatives). | Imperfect detection of species, false negatives |
| Error | It is possible for traces of eDNA from elsewhere to contaminate a sample (false positives). | No analogue in conventional surveys |
| Stage 2 - eDNA lab + bioinformatics pipeline | | Analogues in conventional surveys |
| Species effects | Species differ in extraction efficiency, gene copy number, and PCR amplification efficiency, causing the relationship between input eDNA amount and number of output sequence reads to be species-specific (a 'column' effect). | Species differ in their detectability by human observers. |
| Pipeline effect | PCR stochasticity, normalization steps, and the passing of small aliquots of liquid along the lab pipeline add stochasticity to the total number of output reads per sample replicate (a 'row' effect), including outright detection failure (false negatives). | No analogue in conventional surveys |
| Noise | On top of species and pipeline effects, there is additional noise in the number of reads per species, sample and/or technical replicate (a 'cell' effect). | No analogue in conventional surveys |
| Error | It is possible for traces of eDNA from one sample to contaminate other samples (false positives) or for a sequence to be assigned an incorrect taxonomy (paired false-negative / false-positive errors). | Species identification, error resulting in paired false-negative / false-positive errors |
| Barcoding | Species identification / delineation based on sequence information | Species misidentification resulting in false-positive errors, possibly inducing also an additional false negative, cryptic species |

## Box 2: jSDMs as a tool to model novel community data

In recent years, joint species distribution models (jSDMs) have emerged as the main extension of classical species distribution models for the analysis of community data [60–62]. The key difference between SDMs and jSDMs is that while the former can also model communities, they do so by describing each species individually (stacked SDMs).

A jSDM, however, is a true community model because, additional to the environmental responses of each species, it includes a species-species covariance component, which describes the tendency of species pairs to occur more or less often together than one would expect based on species-specific predictors (species associations, see Fig. B2). Due to their complex likelihood, jSDMs are often challenging to fit, and several numeric strategies, including the latent-variable approximation (e.g. [60]) and Monte-Carlo approximations [96] have been proposed to make these models more scalable to large community data.

The basic jSDM structure can be extended to include additional species correlations via phylogeny or traits, and also spatial predictors. jSDMs can also be extended to fit spatio-temporal data, which allows one to consider additionally asymmetric associations [63,64]. In particular the interpretation of the covariance term has been subject to considerable debate in the field, but we view it now as accepted that species *associations* are not necessarily caused by biotic *interactions* (e.g. [36]; but see [35]). Among other things, this implies that a jSDM will typically not improve the estimation of the fundamental niche [37]. Nevertheless, the ability to partition the community signal into the three classical components of environment, space, and association (Fig. B2), which can further be broken down to sites (communities) and species (i.e., internal structure, see [39] and Fig. B2), provides a rich framework for analyzing spatial community data. Moreover, if some species can be easily observed, conditioning on their presence using jSDMs can also improve predictions [77], which may be relevant for management.

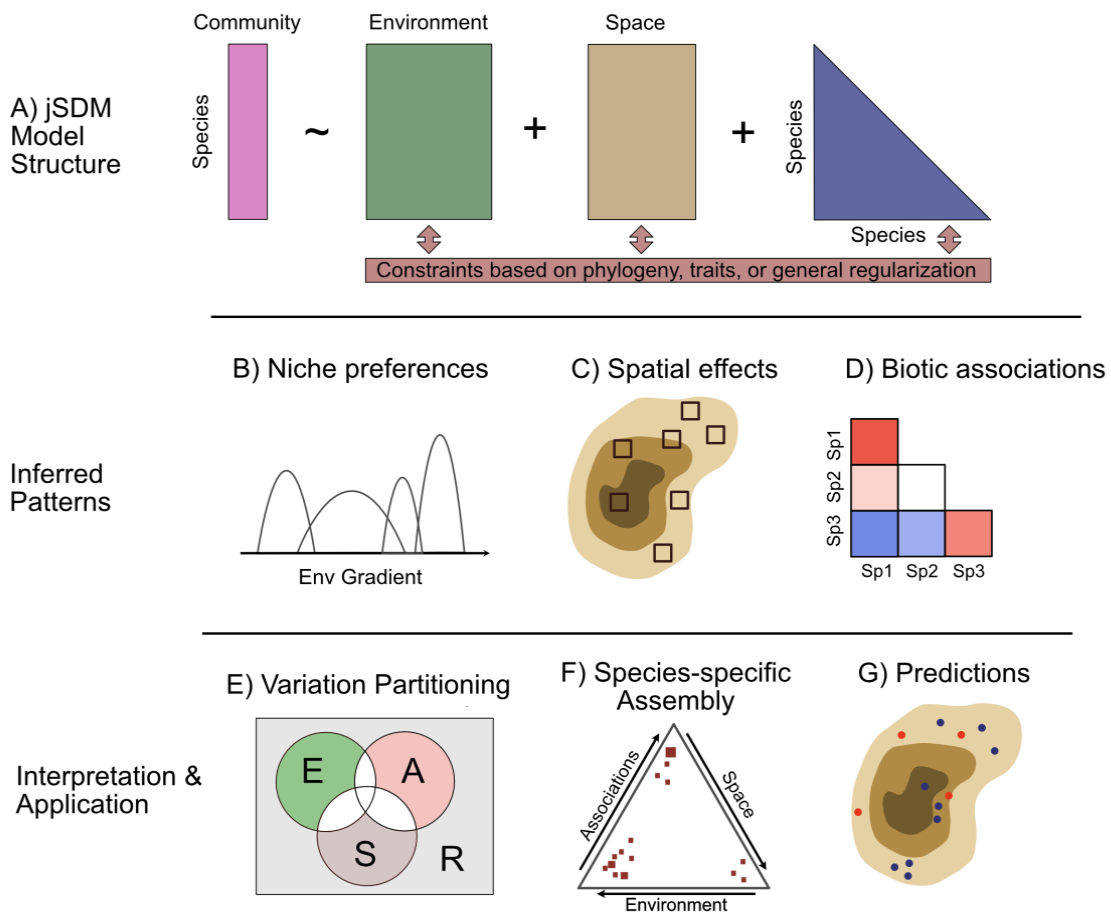

**Fig. B2:** An overview of structure, inferred patterns and interpretation of a jSDM. A) A possible jSDM structure, predicting community composition based on environment, space and species-species covariance. B) Environmental effects show niche preferences C) Spatial effects show spatial clustering of species D) Species-species covariance shows species associations. E) An ANOVA of the entire jSDM (A) can partition community variation into Environment, Space, Associations and Residual components. F) This can further be broken down to species or sites [39], so that we can see the relative importances of the three components to individual species. G) If particular presences are known (red), we can condition on them to improve predictions [77].

## Box 3: An eDNA-enabled biodiversity offset market

One example of institutional redesign enabled by eDNA is the District Licensing Market for the great crested newt (*Triturus cristatus*), a protected species in the UK. Developers are required to survey for the newt when their plans may affect ponds, and to respond to newt detections by paying for mitigation measures. Traditional surveys require at least four visits per pond during the short breeding season, using multiple methods that are only effective at night. Following a study [97] showing that a single eDNA water survey could detect the newt with the same sensitivity as the conventional method (i.e., eDNA detections are *high-quality* and *granular*), the government authorized newt eDNA surveys in 2014, and a private market for eDNA surveys developed to provide the *infrastructure* for *timely* and *trustworthy* information [98].

The switch to eDNA surveys increased survey efficiency, but the the UK's reactive (mitigate after impact) approach was initially left in place. Mitigation measures, such as translocation, can take over a year, with associated costs. In 2018, the UK government took further advantage of eDNA's detection efficiency by implementing an institutional redesign with the District Licensing scheme, in which the ponds across one or more local planning authorities are systematically surveyed with eDNA [99]. The data is then used to fit a species distribution model, which is made into an *understandable* map of discrete risk zones for the newt. Builders can meet their legal obligations at any time by paying for a license, the cost of which depends on the size of their site, a background risk-zone pond occupancy rate, and the number of ponds affected.

The fees from these licenses are mainly used towards the proactive creation and long-term management of compensation habitat including ponds with a one-to-four impact-to-gain ratio. The compensation habitat is directed toward Strategic Opportunity Areas that account for planning-authority building aspirations (*political bargaining*). *Enforcement* is through the same processes that apply to all planning permissions. Both the UK government and a private-public-NGO partnership run versions of District Licensing markets, which together have reported creating hundreds of new ponds and associated habitat. In the future, it would be possible to affect a further institutional redesign by exploiting the multi-species information in the pond water samples to move to multi-species conservation planning and offset markets [100].

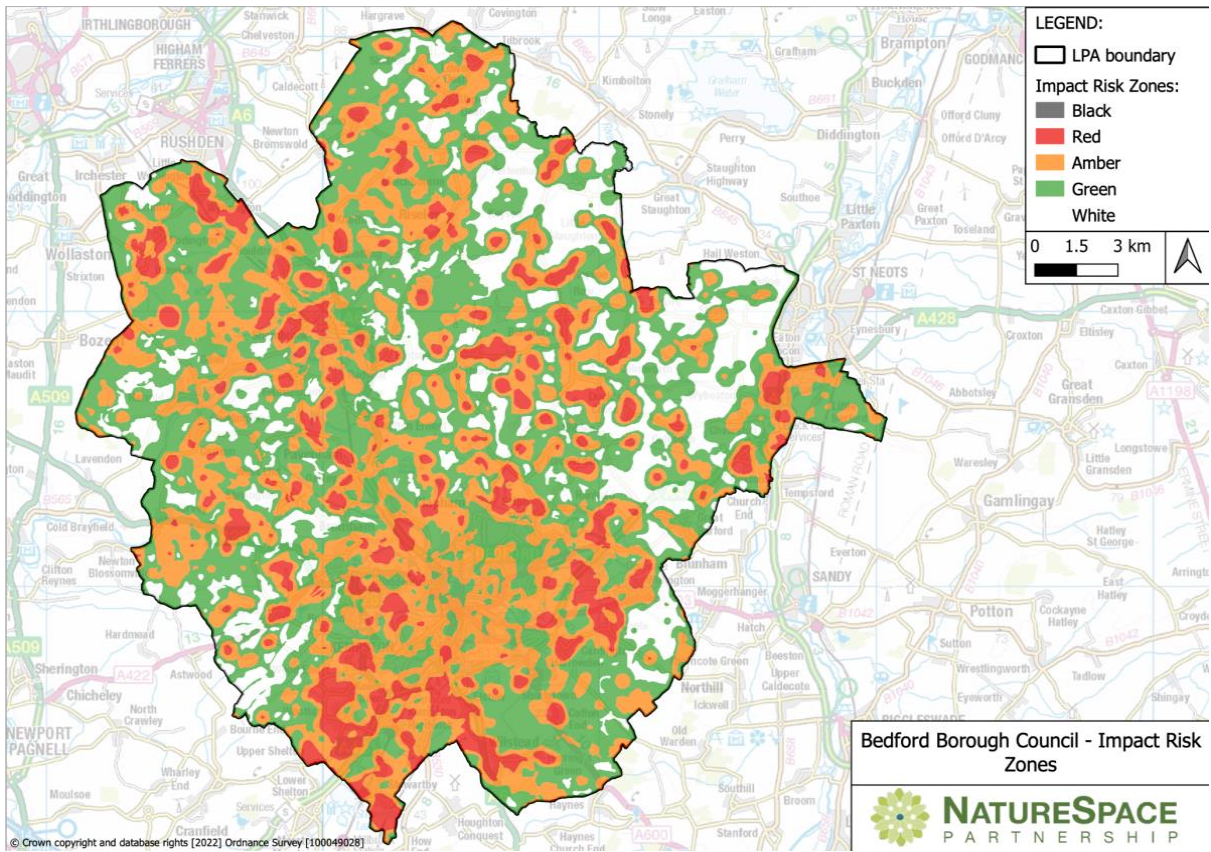

**Fig. B3:** - Risk zone map for great crested newt in one Local Planning Authority (LPA). Reprinted with permission from NatureSpace Partnership.